\documentclass[pra,twocolumn,a4paper,superscriptaddress]{revtex4}

\usepackage{color}

\newif\ifpdf
\ifx\pdfoutput\undefined
\pdffalse 
        \else
        \pdfoutput=1 
        \pdftrue
    \fi
        \ifpdf
          \usepackage[pdftex]{graphicx}
      \else
      \usepackage{graphicx}
      \fi

\newcommand{\tr}{{\textrm{tr}}}

\begin{document}

\title{Detecting entanglement with a thermometer}

\author{Janet Anders\footnote{email JA: janet@qipc.org}}
\affiliation{Quantum Information Technology Lab, Department of Physics,
National University of Singapore, Singapore 117542} 
\affiliation{The School of Physics and Astronomy, University of
Leeds, Leeds LS2 9JT, UK }

\author{Dagomir Kaszlikowski}
\affiliation{Quantum Information Technology Lab, Department of Physics,
National University of Singapore, Singapore 117542} 

\author{Christian Lunkes}
\affiliation{QOLS, Blackett Laboratory, Imperial College, London
SW7 2BZ, UK}

\author{Toshio  Ohshima}
\affiliation{Centre for Quantum Computation, Department of Applied
Mathematics and Theoretical Physics, University of Cambridge,
Wilberforce Road, Cambridge CB3 0WA, UK}
\affiliation{Fujitsu Laboratories of Europe Ltd., Hayes Park Central, Hayes End Road,
Hayes, Middlesex UB4 8FE, UK}

\author{Vlatko Vedral}
\affiliation{The School of Physics and Astronomy, University of
Leeds, Leeds LS2 9JT, UK }

\begin{abstract}
We present a general argument showing that  the temperature as well as other thermodynamical state variables can qualify as entanglement witnesses for spatial entanglement.  This holds for a variety of systems and we exemplify our ideas using a simple free non-interacting Bosonic gas. We find that entanglement  can exist at arbitrarily high temperatures, provided that we can probe smaller and smaller regions of space.  We then discuss the relationship between the occurrence of Bose-Einstein condensation and our conditions for the presence of entanglement and compare the respective critical temperatures.  We close with a short discussion of the idea of seeing entanglement as a macroscopic property in thermodynamical systems and its possible relation to phase transitions  in general.
\end{abstract}

\maketitle

\newpage

\section{Introduction}

Entanglement is a fundamental feature of quantum mechanics and is seen as one of the most important resources in quantum information theory. It provides the key ingredient for teleportation schemes \cite{1,2}, one-way quantum computer \cite{3,4} and many quantum cryptography protocols \cite{5,6}. Furthermore, entanglement may help us understand and explain a variety of quantum phenomena, as for example quantum phase transitions \cite{22}.

The experimental verification of the existence of entanglement in a system under observation is however  still a challenging problem. One solution is to perform a full state tomography to be able to write down the full density matrix describing the state of the system. This is however frequently unfeasible because of the multitude of required measurements. It furthermore raises the question of how to mathematically confirm entanglement once the state is known, or What criterion do we have, to decide whether a known density matrix is entangled?
A second way is to measure an experimentally feasible observable, an \emph{entanglement witness}  (EW) \cite{7,8}.  In contrast to a full state tomography, EWs are designed to detect only one property of the system - its entanglement - without  need to know other details of its state. This approach can reduce the experimental effort greatly. The most prominent example of an EW are the Bell-inequalities - a set of measurements whose outcomes have to obey an inequality relation if the state was separable. The violation of a Bell-inequality tells immediately that the system was entangled, even though one does not know what state is observed.  Entanglement witnesses have been established to observe entanglement in a variety of systems such as qubit systems \cite{7,8}, spin chains \cite{9,10,10b}, and harmonic chains \cite{11,12}. However, to the best of our knowledge, no attempts have been made to use EWs for general continuous variable systems such as the spatial correlations in a Bosonic gas. 

The quintessential problem is to identify entanglement witnesses that are significant for generic systems and which are convenient to measure in practice. Most systems occurring in nature settle to a \emph{thermal equilibrium state}, which can be characterised by macroscopic state variables like the temperature, pressure, and so on. In this paper we argue why entanglement itself can be seen as such a macroscopic property of the system and that it can be observed using easily measurable thermodynamical state variables. Some investigations supporting the way of seeing entanglement as a thermodynamical property have recently appeared \cite{29}; in particular the magnetic susceptibility \cite{13,14} of some solids and their heat capacity \cite{10} were identified as entanglement witnesses. Here we present an argument, that invokes only the Heisenberg principle, to show that separable configurations of a free non-interacting Bosonic gas are macroscopically distinguishable from entangled configuration of the gas. Our approach indicates that the previous findings are not just accidental isolated  instances but that in fact many thermodynamical state variables could serve as witnesses of entanglement. 

This paper is organised as follows. We begin with a short introduction of entanglement witnesses in Section \ref{sec:EWs}. In  Section \ref{sec:formal} we introduce our approach of identifying a thermodynamical state variable as an EW. We exemplify this approach by calculating the lowest energy of a free Bosonic gas in a $d$-dimensional box under the assumption that the gas was in a spatially separable configuration. In Section \ref{sec:transT} we introduce thermal states of the gas and derive the transition temperature at which the gas undergoes a transition from separability to entanglement. We then discuss the relationship between the occurrence of Bose-Einstein condensation and our criterion for the presence of entanglement.  We conclude in Section \ref{sec:conc} by a discussion of our approach, its results and implications, and give a perspective of possible future research.

\section{The concept of entanglement witnesses} \label{sec:EWs}

\begin{figure}[t]
  \begin{center}
   {\includegraphics[width=0.3\textwidth]{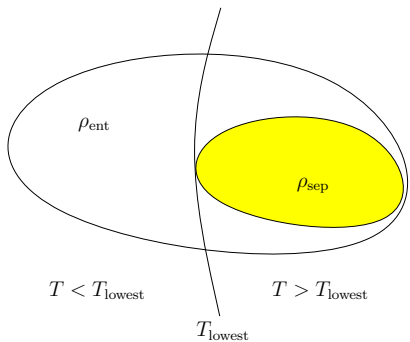}}
    \caption{\label{fig:1} Temperature as an entanglement witness. The figure shows the set of separable states $\rho_{\rm{sep}}$ with respect to some spatial partition and the set of entangled states $\rho_{\rm{ent}}$. We show in this paper that separable states are ``hotter"  than their entangled counterparts. A measurement of the temperature of the system giving a temperature lower than $T_{\rm lowest}$ tells us, that the system under observation was certainly entangled.  The temperature of the system is thus an entanglement witness discriminating entangled states from separable ones. }
       \end{center}
 \end{figure}
 
Entanglement witnesses are observables whose measurement outcomes discriminate all separable states $\rho_{\rm sep}$   (and possibly some entangled states) from a set containing only entangled states $\rho_{\rm ent}$. The idea for the theoretical identification of an observable as an entanglement witness is similar to a {\it reductio ad absurdum} proof. First we assume that the system is in a separable state with respect to a particular partition into two or more subsystems, in our case a partition into $M$ spatial subsystems. Then we derive a property based on this assumption, in our case the minimal energy $E_{lowest}(M)$ of the system. We find a statement of the form\\

\parbox{0.45\textwidth}{If $\rho$ is a separable state w.r.t. the partition into 
$M$ spatial subsystems, then the energy of the system must be at least $E_{lowest}(M)$.\\}
Negating this statement we find that the energy can serve as an entanglement witness for spatial entanglement\\

\parbox{0.45\textwidth}{If the energy of the system is below $E_{lowest}(M)$, $\rho$ is an entangled state w.r.t.  the partition into $M$ subsystems.\\[0.1ex]} 
By measuring a value below the lowest energy $E_{lowest}$ one can tell with certainty that the system under consideration was in an entangled state. When investigating systems in thermal equilibrium, we can define the \emph{temperature} of the system which is usually monotonously related to the thermodynamical internal energy by the equation of state. Thus we can translate the energy $E_{lowest}$ into a temperature $T_{lowest}$ and use the temperature itself as an entanglement witness (see Fig. \ref{fig:1}).
The key observation of our paper is, that we are in fact \emph{able to} construct non-trivial statements of the above form for the expectation values of thermodynamical quantities (for example the energy, temperature or entropy\footnote{We do not want to enter the discussion here of whether one can find an operator representing such thermodynamical state variables. For us it will be sufficient that these variables have the discriminating property mentioned above.}) which  leads to thermodynamical criteria for the occurrence of entanglement.

\section{Method of how to identify a thermodynamical state variable as an EW}\label{sec:formal}

To make a statement of principle we discuss our approach of how to identify a thermodynamical state variable, e.g. the energy, as an entanglement witness on the example of a gas of free non-interacting Bosons. Using thermodynamical relations we can then easily infer bounds for other thermodynamical state variables of the gas such as its temperature. Our argument is based solely on the Heisenberg uncertainty principle which requires, for example for the energy, that there is a non-vanishing zero point energy. The outlined method is hence in principle applicable to all quantum systems; in particular it can be used for more complex, both non-interacting and interacting, Bosonic as well as Fermionic quantum fields. Furthermore, second or higher order observables like the heat capacity can be investigated and used to construct EWs in much the same spirit.  However, whether the method leads to a non-trivial entanglement witness will be up to the specific system and its Hamiltonian.

\subsection{Spatial entanglement}

\begin{figure}[t]
  \begin{center}
  \vspace{1ex}
  
   {\includegraphics[width=0.45\textwidth]{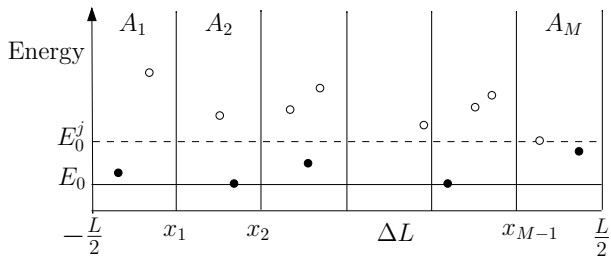}}
    \caption{\label{fig:2a} Schematic picture of a Bosonic gas moving freely in a box of length $L$. The finite size of the box requires that each Boson must carry at least the zero point energy $E_0$. However, if the Bosons are in a spatially separable configuration w.r.t. the partition of space into $M$ subsets $A_j$, $j=1, ..., M$, each Boson has on average  an energy of at least the ground state energy $E_0^j$ in the subset $A_j$. For a free gas, $E_0^j$ is the same for all subsets and $E^j_0 > E_0$ due to the Heisenberg uncertainty relation. The white dots represent Bosons which have energy higher or equal to  $E_0^j$ and thus fulfil Heisenberg in the spatial subsets, whereas the shaded dots lie below this threshold.}
       \end{center}
 \end{figure}

Consider a gas of particles confined in one-dimensional box of length $L$. We want to investigate the entanglement of such systems with respect to a ``spatial partition'' into $M$ subsets $A_j$ of equal size $\Delta L = {L \over M}$, where the spatial subsystems are just the ranges $A_j = (x_{j-1}, x_{j}]$ with $x_j =\left( j - {M \over 2} \right) {\Delta L}$ for $j=1, ..., M$ (see Fig. \ref{fig:2a}).  
Any state $\rho$ describing the system can be written as a classical mixture of a set of pure states $|\chi(i) \rangle$, $\rho = \sum_i p_i \, |\chi(i)\rangle \langle\chi(i)|$, with some probabilities $p_i$ ($\sum_i p_i =1$) that the $i$-th pure state $|\chi(i)\rangle$ was prepared. To be separable with respect to the partition into $M$ spatial subsets means that there exists a decomposition of the state $\rho$ into pure tensor products of each subset, so that\footnote{Here we neglect the symmetrisation of the state for all Bosons. Of the separable states we consider some are symmetric, but not all.  The symmetrised separable states are thus only a subset of the general seperable states and the energy argument derived in the following is in particular true for symmetrised separable states. }

\begin{equation}\label{eq:sep}
\rho_M = \sum_i \, p_i \, |\chi^1(i) \rangle \langle \chi^1 (i) | \otimes ... \otimes  |\chi^M(i) \rangle \langle \chi^M (i)|.
\end{equation}

Let us now calculate the energy of the system, $\langle \hat H \rangle $, under the assumption that the system is in a separable state. Decomposition  (\ref{eq:sep}) tells us, that we can evaluate the energy for each pure tensor product state $|\chi (i) \rangle = \bigotimes_{j=1}^M |\chi^j (i) \rangle$ and later  average these energy contributions with probabilities $p_i$ (In the following we will drop the index $i$ for clearness.). Because of the tensor product structure of $| \chi\rangle$, the average energy is just the sum of the average energies in the subsystems  $A_j$ and since we are interested in a lower bound of the Hamiltonian, let us assume that the system is in the energy minimising state $|\chi^j_0 \rangle$ in each subset $A_j$. The expectation value over the whole system for each pure component $|\chi \rangle$ in (\ref{eq:sep}) is then bounded from below by the sum of the minimal energies in each subset,
\begin{equation}\label{eq:tensor}
\langle \chi|  \hat  H | \chi \rangle^M_{sep} = \sum_{j=1}^M  \langle \chi^j|  \hat  H | \chi^j \rangle \ge \sum_{j=1}^M  \langle \chi_0^j|  \hat  H | \chi_0^j \rangle.   
\end{equation}
Note, that if one would in contrast allow for entangled pure states in the above calculation, inequality  (\ref{eq:tensor}) could be violated and the system's minimal energy could reach values below the sum on the RHS. This would be due to  ``interference'' terms, produced by quantum correlations or entanglement between the Bosons in the different subsets $A_j$.

\begin{figure}[t]
  \begin{center}
  \vspace{1ex}
  
   {\includegraphics[width=0.25\textwidth]{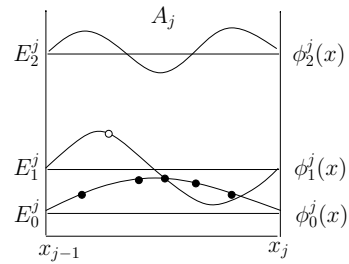}}
    \caption{\label{fig:2b} How to put Bosons in a subset $A_j$ so that the energy is minimised? The ground state for one Boson in subset $A_j = (x_{j-1},x_{j}]$ is denoted by $\phi_0^j(x)$ with energy $E_0^j$, similarly $\phi_1^j$ has energy $E_1^j$ and so on. To minimise the energy, non-interacting Bosons will all accumulate in the ground state (shaded dots) since putting a Boson in a higher excitation e.g. $\phi_1^j(x)$ (white dot) will only raise the energy. }
       \end{center}
 \end{figure}

\subsection{Free non-interacting Bosonic gas}

Up to here the argument is clearly independent of the actual system. The Hamiltonian of a specific system comes in now, when we need the minimal energies in each subsystem $A_j$.  Let us now focus  on a free non-interacting Bosonic gas. We want to put an average number  of $\langle \hat n_j \rangle =  \langle \chi^j_0 | \hat n | \chi^j_0 \rangle$  Bosons in each subset $A_j$, so, as to minimise the energy. The first Boson we put in $A_j$ will occupy the lowest possible one-Boson state $|\phi_0^j\rangle$. We  now want to put another Boson in subsystem $A_j$ raising the energy as little as possible. Since we do not consider a repulsive interaction between the Bosons we clearly  put the second Boson in the same state $|\phi_0^j \rangle$ and similarly for subsequent Bosons (see Fig. \ref{fig:2b}).  The states $| \chi^j_0 \rangle$ minimising the energy for more than one particle in subsystem $A_j$ are  thus just arbitrary  superpositions of number states of the excitation $|\phi^j_0 \rangle$.  If we included  a repulsive interaction between the Bosons, the energy of the system would be raised since either we occupy higher energy excitations or we put more than one Boson in the same state $|\phi_0^j\rangle$ which requires an extra amount of work to overcome the repulsion. Thus the lower bound derived in the following is also valid for repulsively interacting Hamiltonians. The energy of the  states $|\chi_0^j \rangle$ is thus   $\langle \chi^j_0 | \hat H | \chi^j_0 \rangle     \ge  \langle \hat n_j \rangle  ~ \langle \phi^j_0 | \hat H | \phi^j_0 \rangle$, with equality for a non-interacting and inequality for a repulsively interacting gas.

Let us now generalise our derivation (\ref{eq:tensor}) for any mixed $M$-separable state, which can be written as in (\ref{eq:sep}) where each state $| \chi(i) \rangle$ satisfies  $\langle \chi(i)| \hat H| \chi(i) \rangle_{sep}^M  \ge  \sum_{j=1}^M  \langle \hat n_j (i) \rangle  ~ \langle \phi^j_0 | \hat H | \phi^j_0 \rangle$, with $ \langle \hat n_j (i) \rangle$ being the number of Bosons in the $i$-th state in the $j$-th subset. Clearly, the lower bound of the energy of any general $M$-separable state is now just the same as for the pure case,
\begin{eqnarray}
	\langle  \hat H \rangle_{sep}^M 
	&=& \tr[\rho_M ~ \hat H] 
	= \sum_{i} p_i  \langle \chi(i) |  \hat  H|\chi(i) \rangle \\
	&\ge& \sum_{j=1}^{M} \langle \hat n_j  \rangle  ~ \langle \phi^j_0 | \hat H | \phi^j_0 \rangle,
\end{eqnarray}
where the average occupation numbers of each subsystem $A_j$ are  $\langle \hat n_j \rangle = \sum_{i} p_i     \langle \hat n_j (i) \rangle$, summing up to an average total particle number $N$.

To evaluate the one-Boson ground state energies $E_0^j = \langle \phi^j_0 | \hat H | \phi^j_0 \rangle$ in each subsystem we use that the Bosons move \emph{freely} in each subset without any external potential. The ground state functions in each subset $A_j$ are now just the first energy eigenfunctions $\phi^j_0(x) = \sqrt{2 \over \Delta L} \sin { \pi  (x-x_j) \over \Delta L}$ of a free particle in a box of length $\Delta L = L/M$, with energies $E_0^j =  {\hbar^2 \pi^2 M^2 \over 2m L^2}$.
For a $d$-dimensional system with volume $V_d =L^d$ and partitioning into $M$ ranges in each dimension so that the total number of subsets is $M^d$, the energy eigenfunctions become the product of the one-dimensional eigenfunctions in each spatial degree of freedom. The kinetic energy in any dimension will hence be the energy of the one-dimensional case  with the dimension as a pre-factor. The final lower bound on the energy of all $M$-separable states of a free non-interacting or repulsively interacting Bosonic gas in $d$ dimensions is thus
\begin{equation}\label{eq:1}
    \langle \hat H  \rangle^M_{sep} \ge  d N  {\hbar^2 \pi^2 M^2 \over 2m L^2} \equiv E_{\rm lowest}(M).
\end{equation}
This lowest energy can be understood heuristically by virtue of the Heisenberg uncertainty principle. When we look at separable configurations we essentially squeeze the Bosons in boxes of volume $ (L/M)^d$. The uncertainty we have about their positions then requires a minimal momentum of  ${\hbar M / 2 L}$ for each dimension and for each of the $N$ Bosons, so that the minimal energy of the gas is proportional to $d N \, \hbar^2 M^2 /  m L^2$. This is in clear agreement with our derived  formula for all separable states.  

Let us now shortly clarify the mind-boggling question of why the Bosonic gas should care about what we call a subset and behave in a separable way w.r.t. to our partitioning. Of course the Bosons are \emph{not  physically} affected by our imaginary partitioning. The mathematical description (\ref{eq:sep}) of all separable configurations $\rho$ of the Bosons however says  that we can \emph{calculate} the average energy of the gas for the artificial pure tensor product states $|\chi (i) \rangle$, which are the basis in decomposition (\ref{eq:sep}). For these tensor product states the Bosons really have to fulfill the Heisenberg uncertainty relation  in every subset. Nevertheless, there are many ways to mix a state $\rho$ and the occurence of $\rho$ in nature does not imply that it has been mixed in the way we used in our mathematical analysis; the same state could very well be mixed from a set of entangled states ($|\chi(i) \rangle \not =$tensor product). The point is that there \emph{exists} a way to mix $\rho$ out of tensor product states as in  (\ref{eq:sep}) and the resulting average energy corresponds to a separable state. This is regardless of whether this state was achieved by averaging over separable micro-configurations or entangled micro-configurations, thermodynamically speaking.  However, since the energy is an additive quantity, a mixture of only entangled states resulting in a separable configuration must contain at least one entangled state that has an energy above $E_{lowest}$, so that the average is still above $E_{lowest}$.  Such high energy entangled states would not be detected by our energy-analysis and this illustrates that EWs do not detect all entangled states. The situation is very similar to the \emph{bound entangled states} which are not identified as entangled by the PPT criterion.

Our derivation resulting in (\ref{eq:1}) shows that the  observable energy is  an entanglement witness discriminating separable from entangled states. Any free non-interacting Bosonic gas having an energy below the bound (\ref{eq:1}) can not be in a separable state w.r.t.  the $M$  spatial partitions, but must pertain to a state that is definitely entangled. The same strategy of deriving an energy bound applies when we investigate Bosonic gases exposed to an external potential and/or showing internal attractive interactions. Including the specific values of these contributions and evaluating the ground state energy in each subsystem one can in principle derive a lower bound for any given Hamiltonian. The resulting bounds will depend on the particular case and will show dependence on important parameters e.g. coupling constants of the interaction. Such results would allow new insights in the effect of interaction forces on the generation or extinction of entanglement, which we will address in further research. 

Our method can equally be applied to Fermions, but the energy may not be such a successful entanglement witness. Here the Pauli principle prevents the Fermions from condensing into the ground state and the minimal energy of separable states is of the order of the energy of the Fermi sea, the entangled ground state of a Fermi gas. Unfortunately the gap between theses energies will thus be much smaller and depend much stronger on the potential and interaction contributions than in the Bosonic case. Whether some thermodynamical state variables can serve as entanglement witnesses \cite{21} is then a question of how precise measurements of the observables can be made.

\section{Thermal states}\label{sec:transT}

In the thermodynamical limit, where the number of particles  $N$ becomes large, the Bosonic gas can be described within the framework of thermodynamics. This allows us to assign specific macroscopic properties to the gas, e.g. a temperature, a pressure and so on.  The question we address here is: Is the gas in a thermal state for a specific given temperature in a spatially separable configuration or in a spatially entangled configuration? Or, Does a  temperature exist,  below which all thermal states must be entangled? 

The thermal state for any specific temperature $T$ is completely  determined by the Hamiltonian of the system, $\rho_T = {e^{-\beta \hat H} / Z}$, where $\beta = 1/k_B T$ is the inverse temperature and $Z= \tr[e^{-\beta \hat H}]$ the partition function. For most systems, the thermal state takes only a pure form when the temperature reaches zero, otherwise $\rho_T$ is a mixture of multiple excitations. This is also true in our case of a free non-interacting Bosonic gas, which condenses into a full Bose-Einstein condensate (BEC) in the limit $T \to 0$. A BEC is described by a macroscopic wave-function with long-range correlations stretching over the whole condensate \cite{19} making it thereby spatially entangled in number.  On the other hand, for $T \to \infty$, $\rho_T$ ``converges'' to the  complete mixture which is separable for any partitioning\footnote{ This convergence is mathematically problematic since the system is infinite-dimensional. Nevertheless, from a physical perspective we would expect that  the gas behaves more and more classically, the higher the temperature rises and that at some finite high temperature all entanglement vanishes. However it is unclear how big the volume \cite{30} of the set of separable states is and there may well not be any separable states at all.}. 

\subsection{Transition temperature for entanglement}

 Combining our previous microscopic analysis of separable configurations of the gas resulting in the energy bound  (\ref{eq:1}) with the macroscopic internal energy of a free Bosonic gas in  $d$-dimensional space \cite{20}, we are now able to derive a  ``transition temperature''  of entanglement,
 \begin{equation}\label{eq:2}
T_{\rm trans}(M) = {2 \pi \hbar^2 \over k_{\rm B} \,m L^2} \left( {NM^2 \pi \over 2 \, \zeta\left(1 + {d / 2} \right)} \right)^{2 / (2+d)},
\end{equation}
where $\zeta$  is Riemann's zeta-function and $m$ the mass of the Bosons.  The temperature $T_{trans}$ marks the transition of the gas from low temperature entanglement to high temperature separability. Below this threshold entanglement must definitively exist in the system's state, whereas above both, entangled and separable configurations can in principle exist. 
This implies that the temperature of the gas is an entanglement witness in its own right; and so are other thermodynamical state variables, e.g. the pressure, as they are closely related to the energy or temperature via the equation of state. In a similar fashion higher order observables like the heat capacity can be investigated and used to construct EWs.

The transition temperature in  (\ref{eq:2}) represents a macroscopic entanglement criterion, which can be checked experimentally, typically by the mean velocity measurements of particles in the gas. Here we refer to these techniques simply as  ``a thermometer'', thereby motivating our title. Moreover, our derived formula allows us to make general statements about the conditions sufficient for entanglement to occur. In particular, since the zeta-function in  (\ref{eq:2}) takes finite values for all dimensions  $d > 0$, $T_{\rm trans}$ does not vanish for any $d$ and entanglement can exist irrespective of the dimension of the system. From intuition we expect, that a higher number of partitions implies that there exist more states that are entangled with respect to these finer and finer subsets. Indeed the transition temperature in formula  (\ref{eq:2}) grows with the number of partitions  $M$ thus extending the temperature range where only entangled states are present. By making $M$ larger we find that entanglement can in principle exist at arbitrarily high temperatures provided that we can divide space into arbitrarily small parts.  Even if we assume that we can divide space only down to the Planck length, the temperature bound is still very high; of the order of the Planck temperature $T_P \approx 10^{32} K$ if we assume three dimensions and the mass and mass density of the respective Planck units. Moreover, we observe that the transition temperature decreases with growing mass of the particles. So, heavier particles are more difficult to entangle. This is in accordance with the usual argument that more massive particles are less ``quantum'' because of their smaller de Broglie wavelength. Finally, in the classical limit  $\hbar \to 0$  the transition temperature tends to zero and the set of entangled states detected by a temperature measurement vanishes. These properties comfortably confirm the fact that there is no entanglement in the classical world.  The physical significance of the transition temperature becomes even clearer when we contrast it with the critical temperature of Bose-Einstein condensation. 

\subsection{Relation to Bose-Einstein condensation}\label{sec:BEC}

Bose-Einstein condensation \cite{15,16, 17, 18} is a low temperature effect occurring when a significant fraction of a Bosonic gas condenses into the (same) ground state. In our picture we can exclude any BEC in any of the subsystems roughly when we divide space into as many subsets as the number of particles, $M^d = N$, so that each Boson can effectively live in a small box by itself, completely disentangled from all others.  Assuming a fixed particle density $\rho = {N / V_d}$, the transition temperature for entanglement becomes
 \begin{equation}\label{eq:3}
T_{\rm trans}(M^d =N) = {2 \pi \hbar^2 \rho^{2 / d}\over k_{\rm B} \,m} \left( {\pi \over 2 \, \zeta\left(1 +{d / 2}\right)} \right)^{2 / (2+d)}.
\end{equation}
The transition temperature for entanglement $T_{\rm trans}(M^d =N)$  has a similar structure as the BEC critical temperature \cite{20} , $T_{\rm crit}= {2 \pi \hbar^2 \rho^{2 / d}/ k_{\rm B} \,m  \, \left( \zeta\left({d / 2} \right) \right)^{2 / d}}$, having the same dependency on the particle density $\rho$ and the mass $m$. Note however, that the BEC temperature in one- and two-dimensional systems tends to zero with $\zeta\left({d / 2} \right) \to \infty$ and BEC cannot occur at finite temperatures\footnote{This is true for the thermodynamical limit of $N \to \infty$. For finite $N$, BEC can occur as has been pointed out in \cite{20b}.}. In contrast, entanglement can occur  in lower dimensions as $\zeta\left(1 +{d / 2}\right)$ is always finite for $d>0$. This is not surprising since the existence of entanglement is a much weaker property than the occurrence of BEC. However, in three dimensions, the transition temperature for entanglement differs from the critical BEC temperature only by a factor of (approximately) 2. So the gas is entangled already at $T_{trans}$ but for BEC to occur we have to still cool the system down to half the transition temperature.

In the Bose-Einstein condensation experiments with Sodium in Ketterle's group \cite{17}, the number of particles is  $N = 7 \cdot 10^5$ and the size of the confining three-dimensional box is $L = 10\mu {\rm m}$. The researchers observe the first signs of BEC when the temperature reaches  $T = 2 \cdot 10^{-5} K$  which corresponds to a transition temperature $T_{\rm trans}(M)$  for $M=185$ . Comparing  $M$  and  $N$, we see that this agrees roughly with our prediction for entanglement since $M^3 \approx N$.  The range over which the spatially entangled states stretch is then maximally $\Delta L = {L / M} = 5 \cdot 10^{-8} {\rm m}$, being a bit less than the average distance between the particles. So this range, or ``entanglement length'', stretches most likely over only one Boson and no entanglement can be generated. Only when two or more Bosons come closer to each other than the average distance, in our picture there must be two Bosons in one subset at least, they can create entanglement and start to form a small Bose-Einstein condensate. At the temperatures reported by the researchers this situation starts to become more likely and they observe the phase transition into a BE condensate exactly then.

\section{Conclusions}\label{sec:conc}

We have discussed a method to investigate the spatial entanglement of continuous fields on the example of a free non-interacting Bosonic gas. Our method combines two points of view on the same picture. One view is the microscopic one, where we distinguish between spatially separable and entangled configurations of a Bosonic gas in a box. The other is a macroscopic  or  thermodynamical point of view describing the system as a whole without reference to its internal structure. These two views seem not to fit together at first sight. However, thermodynamics itself is the theory of macroscopic properties of a system resulting from its possible micro-configurations. This is the essence of one of the most astonishing formulae in physics, $S = k_B \ln W$,  formulated by Boltzmann  in 1877. 
Following this spirit, our entanglement analysis shows that it is possible indeed to combine the microscopic entanglement view and the macroscopic thermodynamical view to obtain very reasonable predictions. In particular, this  thermodynamical way of looking at entanglement allows us to approximate the BEC critical temperature in three dimensions. This is possible because entangled states apparently form a macroscopic configuration which can be distinguished from the  macro-configuration pertaining to separable states. Our derivation therefore suggests that entanglement can be considered like any other thermodynamical variable: a macroscopic property of the system, which can be observed using easily measurable thermodynamical state variables such as the internal energy and temperature (see also \cite{27,28}). 

Regarding entanglement as a thermodynamical property of the system does not only give us experimentally feasible observables at hand, but it is a necessary prerequisite for the occurrence of Bose-Einstein condensation. The transition of a free Bosonic gas towards a BEC at the critical temperature  $T_{\rm crit}$ is a first order phase transition phenomenon.  But it is not completely inconceivable that the point where separability turns into entanglement is, in fact, a much better indicator of phase transitions in general \cite{22,23}.  Phase transitions, very loosely speaking, occur when many particles produce an effect that is different from the microscopic properties of the constituents; in other words, the particle identity is then lost and submerged into the whole system. Exactly this happens when we cool the Bose gas below the transition temperature and correlations start to stretch over more than one subsystem, thus turning separable states into entangled ones. The key difference between entanglement and condensation we observed is, that entanglement can exist in low dimensions while condensation cannot. How far this way of looking at critical phenomena can be extended and applied to strongly correlated exotic superconductors \cite{24,25,26}, for example, can of course only be clarified by further research.

\acknowledgements
The authors thank M. Wie\'sniak and A. Ekert for insightful discussions. C.L., V.V. and T.O. thank the National University of Singapore for the kind hospitality during their visits.  J.A. acknowledges support of the Gottlieb Daimler und Karl Benz-Stiftung. V.V. thanks the Engineering and Physical Sciences Research Council in UK and the European Union for financial support. This work was supported in part by the Singapore A*STAR Temasek Grant No. 012-104-0040.

\end{document}